\documentstyle[preprint,aps]{revtex}
\newcommand{\be}{\begin{equation}}
\newcommand{\ee}{\end{equation}}
\newcommand{\eq}{\begin{equation}}
\newcommand{\en}{\end{equation}}
\newcommand{\bc}{\begin{center}}
\newcommand{\ec}{\end{center}}

\newcommand{\lsim}{\raisebox{0.3mm}{\em $\, <$}
\hspace{-3.3mm} \raisebox{-1.8mm}{\em $\sim \,$}}
\newcommand{\gsim}{\raisebox{0.3mm}{\em $\, >$}
\hspace{-3.3mm} \raisebox{-1.8mm}{\em $\sim \,$}}
\begin{document}
\draft
\preprint{}
\title{Possible Implications of the Atmospheric, the Bugey, and the Los Alamos
Neutrino Experiments
 \thanks{Work supported in part by Grant-in-Aid for Scientific Research of
the Ministry of Education, Science and Culture \#0560355.}}
\author{Hisakazu Minakata
 \thanks{Electronic address: minakata@phys.metro-u.ac.jp}}
\address{\it Department of Physics, Tokyo Metropolitan University,\\
          1-1 Minami-Osawa, Hachioji, Tokyo 192-03 Japan}
\preprint{
\parbox{5cm}{
TMUP-HEL-9502\\
March 15, 1995\\
Revised: May 31, 1995\\
hep-ph/9503417
}}
\maketitle
\begin{abstract}

A combined analysis of the terrestrial neutrino experiments and the
Kamiokande observation of atmospheric neutrino anomaly is performed
under the assumption of the existence of dark-matter-mass neutrinos,
as suggested by the recent Los Alamos experiment. In the three-flavor
mixing scheme of neutrinos it is shown that the constraints from these
experiments are so strong that the patterns of mass hierarchy and
flavor mixing of neutrinos are determined almost uniquely depending
upon the interpretation of the atmospheric neutrino anomaly.
\end{abstract}
\newpage


There have been accumulating indirect evidences for nonvanishing
masses and the flavor mixings of neutrinos. They include the solar
neutrino deficit\cite{Cleveland} which may be interpreted by either
the Mikheyev-Smirnov-Wolfenstein mechanism\cite{MSW} or the vacuum
neutrino oscillation\cite{BPW}, both being based upon the notion of flavor
mixing. The second in the list is the atmospheric neutrino anomaly first
observed by the Kamiokande experiment\cite{Hirata} and subsequently
confirmed by other detectors\cite{Casper,Goodman}, which strongly
indicates the large-angle flavor mixing of neutrinos.

The recent announcement of the discovery of nonzero neutrino mass by
the Liquid Scintillator Neutrino Detector (LSND) experiment \cite{NYT,Louis}
in Los Alamos may have brought the first direct evidence for the neutrino
masses and the flavor mixing. The experiment may have observed the
neutrino oscillation $\bar{\nu_\mu} \rightarrow \bar{\nu_e}$ with
oscillation parameters
$\Delta m^2 \simeq 1-12 $ eV$^2$ and $\sin ^2 2\theta \simeq 10^{-2} -
10^{-2.5}$, if interpreted by the two-flavor mixing scheme.
The result may be marginally compatible with the earlier results
obtained by the Los Alamos\cite{Durken} and the BNL experiments
\cite{Boro} and by the KARMEN collaboration experiment \cite{Arm}.
Clearly the result, if confirmed by the continuing runs, has tremendous
implications to particle physics and cosmology\cite{PHKC,FPQ,RS}.

In this paper we try to extract the implications of the possible
existence of the dark-matter-mass neutrinos, as suggested by the LSND
result, in the light of the experimental informations from
the underground, the reactor and the accelerator experiments. We first
observe, as many authors do \cite{PHKC,FPQ,RS}, that one cannot
explain the above three phenomena simultaneously by the three-flavor
mixing scheme without introducing sterile neutrinos.
It is simply due to the fact that the three-flavor scheme
cannot accommodate three hierarchically different mass scales,
$\Delta m^2 \simeq (1-12)$eV$^2$ for LSND, $\Delta m^2 \simeq 10^{-2}$eV$^2$
for the atmospheric neutrino anomaly, and
$\Delta m^2 \simeq 10^{-6} - 10^{-5}$eV$^2$ ($\simeq 10^{-10}$eV$^2$) for the
MSW (vacuum mixing) solution of the solar neutrino problem.

We derive the constraints imposed on neutrino masses and mixing angles
via a combined analysis of the reactor and the accelerator data and
the atmospheric neutrino anomaly under the assumption that at least
one of the neutrinos has a mass which falls into the mass range 1-10 eV
which is appropriate for cosmological hot dark matter. This assumption will
be referred to as the assumption of dark-matter-mass neutrinos (DMMN)
hereafter. We employ the mixing scheme based on three-generation neutrinos,
as beautifully confirmed by the LEP experiments\cite{LEP}. It will be
demonstrated that it is essential to use the three-flavor mixing scheme,
rather than optional use of various two-flavor mixings, for drawing
correct interpretation of the data. We will also consider the restrictions
imposed by the neutrinoless double $\beta$ decay \cite{Moe}.

Amazingly, the constraints imposed by a minimal set of data, the
atmospheric and the Bugey \cite{Achkar} experiments, and the assumption of
DMMN are so restrictive as to determine the masses and the mixing patterns
of three flavor neutrinos. Only a few patterns are allowed:
(A) light ``$\nu_e$'' and almost degenerate strongly mixed heavy ``$\nu_\mu$''
and ``$\nu_\tau$'', and its mass-inverted version, or
(B) light ``$\nu_\tau$'' and almost degenerate strongly mixed heavy
``$\nu_e$'' and ``$\nu_\mu$'', and its mass-inverted one.
The choice of the solutions (A) or (B) is dictated by the interpretation
of the atmospheric neutrino anomaly. The pattern (A) follows if we
interpret the atmospheric neutrino anomaly as due to the
$\nu_\mu \rightarrow \nu_\tau$ oscillation, while (B) results if it is
due to the $\nu_\mu \rightarrow \nu_e$ oscillation.

We stress that neither combinations with atmospheric nor with solar
neutrino data are compelling. The interpretation is somewhat more
involved in the latter case because there still exist
three types of solutions to the solar neutrino problem based on the
neutrino flavor mixing; the small and the large-angle MSW solutions in
addition to the vacuum mixing one. The analysis of this combination
will be presented elsewhere\cite{Mina}. Nonetheless, we should mention
that our theoretical prejudice prefers the case with atmospheric neutrino
data over the other one. It is natural to introduce a sterile neutrinos
to accommodate the third experimental data left over in both cases.
It is, however, difficult to explain the atmospheric neutrino anomaly
by introducing the mixing with sterile neutrinos. In doing so one
encounters the trouble with the light-element nucleosynthesis\cite{SSF}.

We make use of one crucial aspect of the atmospheric neutrino data in
our analysis. Namely, the Kamiokande group recently provided a new data
set called the multi-GeV sample\cite{Fukuda}. They consist of the events
with higher energy, $\gsim$ 1.33 GeV, than the previously reported data.
The important feature of the new data is that, because of the higher energy,
the path-length dependence of the oscillation probability can be probed
by measuring the zenith-angle dependence. It is striking that it can be
perfectly fitted by the neutrino oscillation with mixing parameters
$\Delta m^2 \simeq 10^{-2}$ eV$^2$ and $\sin^2 2\theta \simeq 1$\cite{Fukuda}.
Such quantitative agreement with the zenith-angle dependence is the
strongest support for the neutrino oscillation interpretation of the
atmospheric neutrino anomaly.

We now make three basic observations in view of the data of the LSND and
the Kamiokande atmospheric neutrino experiments.
(1) To have gross deficit in the ratio $(\nu_\mu + \bar{\nu_\mu})/(\nu_e +
\bar{\nu_e})$ we need at least one large mixing angle.
(2) To be consistent with the rate of the oscillation events of the
order of $\sim 5\times 10^{-3}$ or less at least one mixing angle has
to be small.
(3) The feature of the atmospheric neutrino data indicates that the one
of the three $\Delta m_{ij}^2$ is of the order of $\simeq 10^{-2}$eV$^2$.
Notice that we cannot have two $\Delta m_{ij}^2$ of the order of
$10^{-2}$eV$^2$ because it contradicts with the assumption of DMMN,
$\Delta m^2 \gsim 1$eV$^2$.

Based on these observations we classify the hierarchy of the neutrino
masses into the following two types:
\begin{equation}
\mbox{a}  : m_3^2 \approx m_2^2 \gg m_1^2
\hskip 2cm
\mbox{b}  : m_1^2 \gg m_2^2 \approx m_3^2
\label{eqn:hierarchy}
\end{equation}
Here the symbols $\approx$ and $\gg$ imply the differences by
$\sim 10^{-2}$eV$^2$ and $\sim 1-100$eV$^2$, respectively.
Throughout the analysis in this paper the relative magnitude of the
masses connected by $\approx$ does not matter. The other types of mass
hierarchies which are obtained by permuting 1, 2, and 3 will
automatically be taken care of because they merely represent relabeling
the mass eigenstates.

We recollect the basic formula of the oscillation probabilities with
three flavors of neutrinos. We introduce the neutrino mixing matrix
$U$ which relates the flavor- and the mass-eigenstates as
$\nu_\alpha = U_{\alpha i}\nu_i$, where the flavor index
$\alpha$ runs over $e,\mu$ and $\tau$ and the mass-eigenstate index $i$
runs over 1 to 3. We assume the CP invariance in the present analysis.
In this case the mixing matrix $U$ is real and contains only three
angles $\theta_{ij}$.

With the mixing matrix the oscillation probability of neutrinos of energy
$E$ after traversing the distance $L$ can be written as
\begin{eqnarray}
P(\nu_\beta \rightarrow \nu_\alpha) &=& P (\bar{\nu_\beta} \rightarrow
\bar{\nu_\alpha})\nonumber\\
	&=& \delta_{\alpha\beta}
-4\sum_{j>i} U_{\alpha i} U_{\beta i} U_{\alpha j} U_{\beta j}
\sin^2(\frac{\Delta m_{ij}^2 L}{4E}),
\end{eqnarray}
where $\Delta m_{ij}^2 = |m_i^2 - m_j^2|.$
As a convenient parametrization of the matrix $U$ we use the so called
standard form of the Kobayashi-Maskawa matrix, which is now adopted for
the neutrino mixing matrix:

\begin{eqnarray}
U=\left[
\matrix {c_{12}c_{13} & s_{12}c_{13} &   s_{13}\nonumber\\
-s_{12}c_{23}-c_{12}s_{23}s_{13} & c_{12}c_{23}-s_{12}s_{23}s_{13} &
s_{23}c_{13}\nonumber\\
s_{12}s_{23}-c_{12}c_{23}s_{13} & -c_{12}s_{23}-s_{12}c_{23}s_{13} &
c_{23}c_{13}\nonumber\\}
\right],
\label{eqn:matrix}
\end{eqnarray}
where $c_{ij}$ and $s_{ij}$ are the short-hand notations for
$\cos \theta_{ij}$ and $\sin \theta_{ij}$, respectively.
We note that the three real angles can all be made to lie in the first
quadrant by an appropriate redefinition of neutrino phases.

The expressions of the oscillation probability are rather cumbersome
involving many angle factors. Therefore, we shall derive the approximate
formulas by taking into account the mass hierarchies and the experimental
parameters of the three experiments. The oscillation probability which
corresponds to the LSND experiment is approximately given by
\begin{equation}
\label{}
P(\bar{\nu_\mu} \rightarrow \bar{\nu_e})=
4c_{12}^2 c_{13}^2 (s_{12}c_{23} + c_{12}s_{23}s_{13})^2
\sin^2(\frac{\Delta m_{12}^2}{4E}L).
\end{equation}
where two terms with $\Delta m_{12}^2 \approx \Delta m_{13}^2$ (which
differ only by $10^{-2}$eV$^2$) are combined and the term with
$\Delta m_{23}^2$ is ignored.
The former procedure can be neatly done by utilizing the orthogonality
relation of the mixing matrix. The latter approximation is completely
legitimate because the term is smaller than others by factor
$10^{-4}-10^{-8}$ owing to the mass hierarchy
$\Delta m_{23}^2/\Delta m_{12}^2 \simeq 10^{-2}-10^{-4}$.

If the atmospheric neutrino anomaly is attributed to the $\nu_\mu
\rightarrow \nu_\tau$ oscillation the relevant formula is
\begin{eqnarray}
P(\nu_\mu \rightarrow \nu_\tau)
&=& 2(s_{12}c_{23}+c_{12}s_{23}s_{13})^2(s_{12}s_{23}-c_{12}c_{23}s_{13})^2
\nonumber\\
&+& 4c_{23}s_{23}c_{13}^2(c_{12}c_{23}-s_{12}s_{23}s_{13})
(c_{12}s_{23}+s_{12}c_{23}s_{13})
		\sin^2 (\frac{\Delta m_{23}^2 L}{4E}).
\label{eqn:osci2}
\end{eqnarray}
In (\ref{eqn:osci2}) the sine-squared factors with large $\Delta m^2$
of $\gsim$1eV$^2$ are replaced by the average value $\frac{1}{2}$,
which can be justified because of the rapid oscillations; the argument of
the sine is $\sim 10-10^3 (10^4-10^6)$ for L$=10(10^4)$ Km for
$\Delta m^2 = 1-100$eV$^2$ and E$=1$GeV

We note that there exist the possibility that the atmospheric neutrino
anomaly is due to the $\nu_\mu \rightarrow \nu_e$ oscillation, the
possibility one might not naively expect. It is perfectly consistent with
a small event rate in the LSND experiment because the relevant scales of
path length and neutrino energy involved in these two experiments are much
different. In this case the formula for the oscillation probability to be
used is
\begin{eqnarray}
P(\nu_\mu \rightarrow \nu_e)
&=& 2c_{12}^2c_{13}^2(s_{12}c_{23}+c_{12}s_{23}s_{13})^2\nonumber\\
&-& 4s_{12}s_{23}c_{13}^2s_{13}(c_{12}c_{23}-s_{12}s_{23}s_{13})
		\sin^2 (\frac{\Delta m_{23}^2 L}{4E}).
\label{eqn:osci3}
\end{eqnarray}

Finally the formula for the Bugey experiment takes the form
\begin{eqnarray}
1-P(\bar{\nu_e} \rightarrow \bar{\nu_e})
&=& 2c_{12}^2c_{13}^2(1-c_{12}^2c_{13}^2)\nonumber\\
&+& 4s_{12}^2c_{13}^2s_{13}^2
		\sin^2 (\frac{\Delta m_{23}^2 L}{4E}),
\label{eqn:Bu}
\end{eqnarray}
where the terms with $\Delta m_{12}^2$ are averaged as before. It can be
justified because the argument of the sine term is of the order of $10-10^3$
with $\Delta m^2 = 1-100$eV$^2$, E$=4$MeV, and L$=40$m, the typical
parameters of the Bugey experiment. The second term of (\ref{eqn:Bu}) may
be neglected (as we will do) because sine-squared factor is $\sim 10^{-2}$
for $\Delta m^2 = 10^{-2}$eV$^2$.

We first examine the case that the atmospheric neutrino anomaly is
attributed to the $\nu_\mu \rightarrow \nu_\tau$ oscillations.
Our discussion does not distinguish the types a and b
until we address the constraint due to the double $\beta$ decay.

We demand, for consistency with the gross features of the LSND, the Bugey,
and the atmospheric neutrino experiments, the following constraints:
\begin{equation}
\label{eqn:LSND}
c_{12}^2c_{13}^2(s_{12}c_{23}+c_{12}s_{23}s_{13})^2
 \equiv \epsilon \lsim 10^{-3}
\end{equation}
\begin{equation}
\label{eqn:Bugey}
c_{12}^2c_{13}^2(s_{12}^2c_{13}^2+s_{13}^2)\lsim \delta
=2.5\times 10^{-2}
\end{equation}
\begin{equation}
\label{eqn:atmo1}
(s_{12}c_{23}+c_{12}s_{23}s_{13})^2(s_{12}s_{23}-c_{12}c_{23}s_{13})^2
\leq 0.1,
\end{equation}
\begin{equation}
\label{eqn:atmo2}
4c_{23}s_{23}c_{13}^2(c_{12}c_{23}-s_{12}s_{23}s_{13})
(c_{12}s_{23}+s_{12}c_{23}s_{13})
\simeq 1.
\end{equation}
The constraints (\ref{eqn:LSND}) comes from the LSND experiment. We
treat $\epsilon$ as a small number of the order of $\sim10^{-3}$ or less.
Our discussion will be insensitive to the number and we use it as a
tentative guide when we address the consistency with other experiments.
The equation (\ref{eqn:Bugey}) is due to the bound
$1-P(\bar{\nu_e} \rightarrow \bar{\nu_e}) \lsim 5$ \%
obtained in the Bugey experiment\cite{Achkar}. It includes statistical
and systematic uncertainties. The remaining two restrictions are from
the features of the atmospheric neutrino data that the zenith-angle
dependence is well described by an effective two-flavor-mixing ansatz
with $\Delta m^2 \simeq 10^{-2}$eV$^2$ and $\sin^2 2\theta \simeq 1$.
The constraint (\ref{eqn:atmo1}) arises from a mild requirement that
the first term of (\ref{eqn:osci2}) should be less than 0.2 so as not
to disturb the effective two-flavor description. We emphasize that
the constraints from the atmospheric neutrino data take the simple
forms (\ref{eqn:atmo1}) and (\ref{eqn:atmo2}) because of the mass
hierarchy $\Delta m_{12}^2 \approx \Delta m_{13}^2 \gg \Delta m_{23}^2$.

We first notice that from the Bugey constraint(\ref{eqn:Bugey})
that $X\equiv c_{12}^2c_{13}^2$ must satisfy the inequality
$X^2 - X + \delta \geq 0$. This inequality is so powerful
that restricts the value of $X$ into the two tiny regions
$0 \leq X \leq  \delta$ and $1-\delta \leq X \leq 1$.
On the other hand, we must have $c_{13}^2 \simeq O(1)$ in order to
satisfy the requirement (\ref{eqn:atmo2}). Thus, we have either
$c_{12}^2 \simeq \delta$ or $c_{12}^2 \simeq 1$ corresponding to
the small-$X$ and the large-$X$ solutions, respectively.
It is also required that $c_{23}s_{23} \simeq \frac{1}{2}$ in order
to maximize (\ref{eqn:atmo2}). The small-$X$ solution is then
inconsistent with (\ref{eqn:atmo2}). We end up with the unique solution
\begin{equation}
\label{eqn:solA}
(\mbox{A}) \hskip 1cm s_{12}^2 \approx s_{13}^2 \simeq \epsilon, \hskip 1cm
c_{23}^2 \approx s_{23}^2 \simeq \frac{1}{2}
\end{equation}
where we have also utilized the LSND constraint (\ref{eqn:LSND}) to push
$s_{12}^2 \simeq \delta$ down to $s_{12}^2 \simeq \epsilon \lsim10^{-3}$.

We have explicitly verified that the allowed mixing pattern implied by
(\ref{eqn:solA}) is physically unique throughout the varying mass
hierarchies obtained by the cyclic permutations of 1-3 of
(\ref{eqn:hierarchy}), as it should be.
Namely, the light ``$\nu_e$'' and the almost degenerate strongly mixed
heavy ``$\nu_\mu$'' and ``$\nu_\tau$'' for the type-a, and the heavy
``$\nu_e$'' and the almost degenerate strongly mixed light ``$\nu_\mu$''
and ``$\nu_\tau$'' for the type-b cases.

We now turn to the case that the atmospheric neutrino anomaly is
caused by the $\nu_\mu \rightarrow \nu_e$ oscillation. In this
case we replace the requirements (\ref{eqn:atmo1})
and (\ref{eqn:atmo2}) by
\begin{equation}
c_{12}^2c_{13}^2(s_{12}c_{23}+c_{12}s_{23}s_{13})^2 \leq 0.1,
\label{eqn:atmo3}
\end{equation}
and
\begin{equation}
-4s_{12}s_{23}c_{13}^2s_{13}(c_{12}c_{23}-s_{12}s_{23}s_{13}) \simeq 1,
\label{eqn:atmo4}
\end{equation}
respectively.
By the similar procedure one can show that the consistent solution
of the requirements (\ref{eqn:LSND}), (\ref{eqn:Bugey}), (\ref{eqn:atmo3}),
and (\ref{eqn:atmo4}) is uniquely given by
\begin{equation}
(\mbox{B}) \hskip 1cm c_{12}^2 \approx c_{23}^2 \simeq \sqrt{\epsilon},
\hskip 1cm c_{13}^2 \approx s_{13}^2 \simeq \frac{1}{2}.
\end{equation}
The solutions of the other type of mass hierarchies can be obtained by
the similar manner and correspond to the redefinition of the mass eigenstates.
The allowed mixing pattern is again physically unique:
The light ``$\nu_\tau$'' and the almost degenerate strongly mixed
heavy ``$\nu_e$'' and ``$\nu_\mu$'' for the type-a, and the heavy
``$\nu_\tau$'' and the almost degenerate strongly mixed light
``$\nu_e$'' and ``$\nu_\mu$'' for the type-b mass hierarchies.

We note that the solutions (A) and (B) are subject to the additional
constraints from other terrestrial experiments. While the solution (A)
solves them automatically the nontrivial constraints arise for (B).
In particular, the most stringent one comes from the Fermilab E531
experiment \cite{E531} for $\Delta m^2 \gsim 3$eV$^2$ and the $\nu_\mu$
disappearance experiment by the CDHS group \cite{CDHS} for
$\Delta m^2 \lsim 3$eV$^2$. If we take the rate of the appearance events
reported in \cite{LSND} at its face value the solution (B) may be excluded
apart from tiny regions. To establish the rate, however, an additional
run of the experiment as well as its careful analysis would be required.

In the case of Majorana neutrinos a further constraint emerges
from the non-observation of the neutrinoless double $\beta$ decay.
The quantity

\begin{equation}
<m_{\nu e}> = \sum_{j=1}^{3} \eta_j |U_{ej}|^2 m_j
\end{equation}
is constrained to be less than $\sim$1eV by the experiments\cite{Moe}
where $\eta_j =  \pm 1$ is the CP phase. Notice that we are working
with the representation in which the mixing matrix is real under the
assumption of CP invariance.

Generally speaking, the constraint from the double $\beta$ decay
distinguishes between the type-a and the type-b mass hierarchies. In
the type-a case there is a chance for cancellation between nearly
degenerate two masses, but no chance in the type-b case
because the heavy mass is carried by a unique mass eigenstate.

New features, however, arise in our consistent solutions obtained
above. We first discuss the case of atmospheric neutrino anomaly
due to the $\nu_\mu \rightarrow \nu_\tau$ oscillation.
It can be shown that in the type-a mass pattern the double $\beta$
constraint is automatically satisfied because the heavy masses are always
multipled by small angle factors. On the contrary, the angle factors in
front of the unique heavy mass are always of the order of unity in the
type-b mass hierarchy. Therefore, there is no consistent solution of the
double $\beta$-decay constraint for Majorana neutrinos in the type-b
hierarchy.

In the case of atmospheric neutrino anomaly due to the $\nu_\mu
\rightarrow \nu_e$ oscillation, the situation is somewhat different.
In the type-a mass pattern there is a trouble because almost
degenerate heavy masses are multiplied by $O(1)$ coefficients and a
tuning, i.e., $s_{12}^2c_{13}^2 = s_{13}^2$ to better than 0.1, is
required for cancellation in addition to the requirement of opposite
CP parities. On the contrary, in the type-b hierarchy, there is no
trouble with the double $\beta$-decay constraint because the heavy mass
is multiplied by small coefficients of the order of $\sqrt{\epsilon}$.

Thus, we have shown in this paper that the neutrino masses and the mixings
are strongly constrained by the atmospheric and the terrestrial experiments
under the assumption of DMMN suggested by the LSND experiments. The
constraint is so severe that the mass and the mixing patterns are
determined almost uniquely within the uncertainties of the neutrino
types and the interpretations of the atmospheric neutrino anomaly. In
the case of the Majorana neutrinos the additional constraint from the
double $\beta$ decay selects out the unique natural solution in each
interpretation.

Finally we give a few remarks.

\noindent
(1) Our analysis in this paper is less powerful in constraining
the absolute values of the neutrino masses than in restricting the relative
masses and the mixing angles. All the constraints would be cleared by,
for example, the type-a solution with $m_1 =$6 eV, $m_2 =$6.5 eV,
and $m_3 = (6.5+\epsilon)$ eV, which is also consistent with the direct
measurements \cite{Otten} and the cosmological considerations\cite{Kolb}.

\noindent
(2) We have not performed a full three-flavor analysis of the atmospheric
neutrino anomaly but relied on the effective two-flavor interpretation of
either $\nu_\mu \rightarrow \nu_\tau$ or $\nu_\mu \rightarrow \nu_e$
channels. While the point deserves further study we are under a strong
feeling that coexistence of the two channels does not spoil our
solutions obtained in this paper. They certainly survive in the case of
equal contributions of these two channels.

This work is supported in part by Grant-in-Aid for Sciectific Research of the
Ministry of Education, Science and Culture, \#0560355.

Note Added: After submitting the earlier version of this paper we became
aware of two reports from the LSND group \cite{LSND,Hill} with mutually
conflicting conclusions.

\end{document}